\newcommand{\beqn}{\begin{equation}}
\newcommand{\eeqn}{\end{equation}}

\documentclass[aps, pra, reprint, groupedaddress]{revtex4-2}

\usepackage{dcolumn}
\usepackage{bm}

\usepackage{epsfig}
\usepackage{graphicx}
\usepackage{amssymb}
\usepackage{amsmath}

\begin{document}

\title{Diode laser spectroscopy of methyl iodide at 850 nm}

\author{A. Lucchesini}
\email{lucchesini@ino.cnr.it}
\affiliation{Istituto Nazionale di Ottica - CNR - S.S. ``Adriano Gozzini'' \\
Area della Ricerca - Pisa - Italy}

\date{\today}


\begin{abstract}
 Using the Tunable Diode Laser Absorption Spectroscopy (TDLAS) 82 CH$_3$I ro-vibrational overtone absorption were detected for the first time between 11660 and 11840 cm$^{-1}$ (844 -- 857 nm), with strengths estimated to be around $10^{-27}$ -- 10$^{-26}$ cm/molecule.
The lines have been measured utilizing commercial heterostructure F--P type diode lasers, multipass cells and the wavelength modulation spectroscopy with second harmonic detection technique. A high modulation amplitude approach was adopted for the analysis of the line shapes. Self-broadening coefficients have been obtained for two lines.
\end{abstract}
\maketitle

\section{Introduction}

This work on methyl iodide or iodomethane (CH$_3$I) is the natural continuation of the previous ones on methyl halides CH$_3$F~\cite{ref:lucch2013} and CH$_3$Cl~\cite{ref:lucch2016} (C$_{3\upsilon}$ symmetry group), detected at 850 nm in the gas phase by the author. In fact, their absorption bands in the infrared (IR) and in the near-infrared (NIR) part of the e.m. spectra are very similar.

Methyl iodide is a prolate symmetric top molecule, which is used in agriculture as a pesticide and is present in the earth's atmosphere, classified as halogenated volatile organic compound (HVOC). It participates to the ozone layer depletion~\cite{ref:laube2022}. This molecule is one of the most investigated in the IR, where it can be finely studied by spectroscopy based on semiconductor sources.

The CH$_3$I absorption spectrum between 850 nm and 2.5 $\mu$m has been detected in the distant past by Gehrard and Luise Herzberg~\cite{ref:herz1949} using an infrared prism spectrometer, when they classified the observed overtone and combination bands.
Even older is the work of Verleger on methyl halides at wavelengths below 1.2 $\mu$m~\cite{ref:verle1935}, where overtone bands have been observed on photographic plates through a 3 m grating monochromator.
Methyl iodide optical absorption has been more recently studied in the NIR by Ishibashi and Sasada~\cite{ref:Ishi2000} with diode lasers as the sources to detect the 2$\nu_4$ overtone band at 6,050 cm$^{-1}$ with sub-Doppler resolution in a Fabry--Perot (F.--P.) type cavity measurement cell.

In this experimental work a tunable diode laser spectrometer (TDLS) with the frequency modulation and the second harmonic detection technique was used to observe the CH$_3$I ro-vibrational band at around 11740 cm$^{-1}$ (850 nm) with a resolution of 0.01 cm$^{-1}$.
This NIR band is presumably related to the third overtone of the $\nu_1$ quanta of C--H stretching excitation~\cite{ref:herz1949} or a combination of overtones, such as, for instance, 2$\nu_1$ + 2$\nu_4$. This can lead to overlapping absorption bands, whose upper states can be coupled through Fermi and Coriolis resonances~\cite{ref:law1999}.
Therefore it is difficult to identify the right ro-vibrational quanta, unless sophisticated techniques are employed, such as for example microwave-optical double resonance~\cite{ref:Ishi2001}.

This absorption band is very weak, as the dipole moment is even less than the one of CH$_3$F and CH$_3$Cl, as the C--I bond length is shorter the C--F and C--Cl ones, therefore lower absorption strength for this molecule is expected. It is therefore necessary to use long optical paths and noise reduction techniques to observe and study these absorption lines.

Frequency Modulation Spectroscopy (FMS)~\cite{ref:bjor1980}, conventionally called Wavelength Modulation Spectroscopy (WMS) when the modulation frequency is much lower than the sampled line-width, was used on this occasion as a noise-reduction technique.

\section{Experiment}
\subsection{Experimental setup}
\label{exp}

\begin{figure}
\resizebox{\columnwidth}{!}{
  \includegraphics{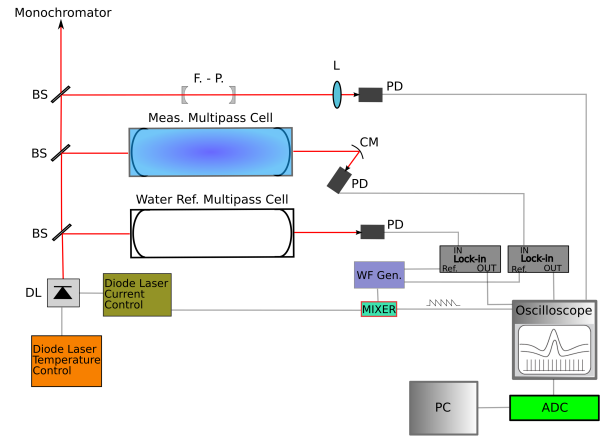}
}
\caption{Outline of the experimental apparatus. ADC: analog-to-digital converter; BS: beam splitter; CM: concave mirror; DL: diode laser; F. - P.: confocal Fabry-Perot interferometer; L: lens; PC: desk-top computer; PD: photodiode; WF Gen.: waveform generator.}
\label{ExperApp}
\end{figure}
The schematic of the experimental apparatus is illustrated in Fig.~\ref{ExperApp} and follows the previous work on CH$_3$Cl~\cite{ref:lucch2016}.
The employed source was the single mode AlGaAs/GaAs double heterostructure F--P type diode laser (DL), Thorlabs L852P100, with maximum power $\simeq 100$ mW cw in {\it free-running} configuration.

The DL temperature control is very critical as its typical emission wavelength varies as about 0.1 nm/K, therefore a customized bipolar temperature controller was adopted to drive a Peltier junction coupled to the DL mount. This guaranteed a stability of $\sim 0.002$ K per hour.
Also the DL current needs precise and fine control, since the characteristic slope of the DL emission is about 0.01 nm/mA. Consequently a low noise DL current controller was needed: for this experiment a custom-made current source was used, operating between 0 and 250 mA, with an accuracy of $\pm\ 2.5$ $\mu$A.
To modulate and sweep the emission source, a sine wave carrier from a low noise waveform generator was mixed with the ramp extracted from the oscilloscope sawtooth signal and then sent to the DL current controller.

Two custom Herriott-type astigmatic multipass cells with optical path lengths of 30 m were used to house the sample and reference gases. The~latter contained water vapor at room temperature (RT) with a partial pressure $\simeq 20$ torr and~was used as the reference gas for wavenumber measurements  (for this purpose, the HITRAN2016 molecular spectroscopic database~\cite{ref:hitran2017} was adopted) and to verify whether the water absorption lines could interfere with the observed CH$_3$I lines.
At the exit of the measurement cell, a concave mirror focused the laser beam on the active spot of the photodiode (PD), with the purpose of reducing the mechanical noise coming from the optical leverage of the spectroscope set.

Pre-amplified silicon PDs (Centronic OSD5-5T, with active surface of 2.52 mm diameter) were used as detectors, whose outputs were sent to lock-in amplifiers tuned to the carrier frequency ($f \sim 5$ kHz) coming from the waveform generator.
It was verified that the residual phase errors in the synchronization of the two lock-in amplifiers could induce at most a corresponding wavenumber error of 0.003 cm$^{-1}$.
A confocal 5 cm F.--P. interferometer (f.s.r. = 0.05 cm$^{-1}$) was utilized to check the DL emission mode and the linearity of its emission frequency, while a 35 cm focal length Czerny--Turner monochromator with 1,180 lines per mm grating was used for the rough wavelength check ($\pm\ 0.01$ nm).

The vacuum in the sample cell was obtained by  a double stage rotary pump with limit pressure $< 1 \times 10^{-4}$ torr, and the pressure inside the cell was directly measured by a capacitive pressure gauge (accuracy $\pm 0.5$ torr). All experiments were conducted at RT and at pressure values ranging from 20 to 90 torr.
Analytical grade methyl iodide (Sigma-Aldrich, 99\% purity, with silver as stabilizer) was used as supplied. The sample was contained in a stainless steel tube and it was allowed to flow into the evacuated measuring cell at its vapor pressure.

\subsection{Wavelength Modulation Spectroscopy}
\label{WMS}
The transmittance $\tau(\nu)$ after the sampled gas can be described by the Lambert-Beer expression:
\begin{equation}
\tau(\nu) =  e^{-\sigma(\nu) \, z}
\label{eq:beer}
\end{equation}
where $z=\rho\,l$ is the product of the absorbing species density $\rho$ (molecule/cm$^3$) and the optical path $l$ (cm) of the radiation through the sample, namely the {\it column amount} (molecule/cm$^2$), and $\sigma(\nu)$ is the absorption cross-section (cm$^2$/molecule), which follows the line-shape behaviour, but in regimes of small optical depth [$\sigma(\nu) z \ll 1$] as in most experimental conditions, Eq.~(\ref{eq:beer}) can be approximated by
\begin{equation}
\tau(\nu) \simeq 1 - \sigma(\nu)\, z \, .
\label{eq:absor}
\end{equation}

The WMS technique was achieved here by modulating the the source emission frequency $\bar{\nu}$ at $\nu_\mathrm{m} = \omega_\mathrm{m}/2\pi$, with amplitude $a$:
\begin{equation} \label{eq:freq}
\nu = \bar{\nu} + a \cos \omega_\mathrm{m} t \, .
\end{equation}

The transmitted intensity could then be written as a cosine Fourier series:
\begin{equation}
\tau(\bar{\nu} + a\, \cos \omega_\mathrm{m} t) =
\sum_{n=0}^\infty  H_n(\bar{\nu}, a) \, \cos n \omega_\mathrm{m} t
\label{eq:mod}
\end{equation}
where $H_n(\bar{\nu})$ is the $n$-th harmonic component of the modulated signal. With a lock-in amplifier tuned to a multiple $n \nu_\mathrm{m}$ ($n = 1, 2, ...$) of the modulation frequency, an output signal proportional to the $n$-th component $H_n(\bar{\nu})$ is obtained. In case the value of $a$ is chosen smaller than the line-width, the $n$-th Fourier component is proportional to the $n$-th derivative of the original signal:
\begin{equation}
H_n(\bar{\nu}, a) = \frac{2^{1-n}}{n!} \, a^n \,
\left. \frac{d^n \tau(\nu)}{d\nu^n}\right|_{\nu=\bar{\nu}} ,
\quad\quad n \ge 1 \, .
\label{eq:comp}
\end{equation}
In order to detect very low absorbances, a high modulation amplitude regime is required, which implies a modulation index $m \equiv a/{\Gamma} \gg 0.1$, where ${\Gamma}$ is the absorption line-width.

Line position measurements were carried on at pressures around 30 torr. In this condition, where the collisional effect was not negligible, the absorption line shape was well described by the Voigt profile, a convolution of the Gaussian (Doppler regime) and Lorentzian (collisional regime) functions:
\begin{equation}
f(\nu)=\int_{-\infty}^{+\infty}{{\exp{[-(t-\nu_{\circ})^2 /
{\Gamma_\mathrm{G}^\mathrm{2}} \ln2]}\over(t-\nu)^2+{\Gamma_\mathrm{L}^\mathrm{2}}}}dt
\label{eq:voigt}
\end{equation}
where $\nu_{\circ}$ \ is the gas resonance frequency, ${\Gamma_\mathrm{G}}$ and ${\Gamma_\mathrm{L}}$ are the Gaussian and the Lorentzian half-widths at half-the-maximum (HWHM), respectively.
Second-order effects, such as speed-changing collision or Dicke narrowing~\cite{ref:dicke1953}, were not observed within the sensitivity of the apparatus and therefore were not taken into account.

The phase detection technique, obtained by tuning the lock-in amplifiers to twice the modulation frequency ($\sim 10$ kHz), produced a line-shape signal that was closer to the 2nd derivative of the absorption feature as the minor was the amplitude of the modulation.
The symmetry of the line shape was not perfect due to the simultaneous amplitude modulation of the source signal as the injection current of the diode laser varied; indeed, this effect becomes negligible in high modulation regimes.
Such ``$2\!f$ detection'' has the advantage of a flat baseline of the signal, but cannot avoid optical interferences, coming from the many reflecting surfaces present in the optical path. This is the main drawback of this spectroscopic technique, which in the specific case has limited the detection sensitivity to absorbances higher than $1 \times 10^{-7}$.
 Under these conditions this technique could not provide a reliable measurement of the intensity parameter when applied to very weak resonances.
For this reason it was possible only to estimate that the strength of the CH$_3$I absorption lines here lies in the interval $1 - 30 \times 10^{-27}\mathrm{cm/molecule}$ from a comparisonwith the absorption line strengths of the water vapor present in the same spectrum, centered around 11740 cm$^{-1}$, where the strongest lines were found.

To obtain line positions and widths with good reliability even for the weakest lines, the values of the modulation index $m$ have been set around 2.0--2.3. This substantially improved the S/N ratio, but did not allow the use of  Eq.~(\ref{eq:comp}) any more.
The approximate function that represents the absorption line distorted by the modulation has been specifically evaluated and is reported in the Appendix; a nonlinear least-squares fit method was applied to extract line parameters from the collected profiles.

Finally, the following expression of the collisional half-width at half-maximum (HWHM) as a function of pressure was adopted for the line broadening calculations:
\begin{equation}
{\Gamma_\mathrm{L}}(p) = \gamma_\mathrm{self} \, p
\end{equation}
where $p$ is the sample gas pressure, and $\gamma_\mathrm{self}$ is the gas self-broadening coefficient.

\section{Experimental results}
\label{exp}
The 82 CH$_3$I absorption lines have been detected and are listed in Table~\ref{tab:listCH3I}, where the maximum error on the wavenumber ($\nu'$) was within the second decimal unit, referring to the ten times more precise H$_2$O atlas~\cite{ref:hitran2017}.
The wavelengths are reported for convenience in air at $T$ = 294 K, following the work of Edl\'en~\cite{ref:edl1966}.
\begin{table}
\caption{\label{tab:listCH3I}Wavenumbers and wavelengths (in air at room temperature) of the detected CH$_3$I absorption lines, with the maximum error within the second decimal unit.}
\begin{ruledtabular}
\begin{tabular}{cccc}
$\nu'$ (cm$^{-1}$) & $\lambda$ (\AA) & $\nu'$ (cm$^{-1}$) & $\lambda$ (\AA) \\
\noalign{\smallskip} \hline \noalign{\smallskip}
11659.92&8574.08&11697.76&8546.35 \\
11660.13&8573.93&11698.02&8546.16 \\
11664.91&8570.41&11700.55&8544.31 \\
11665.03&8570.33&11700.85&8544.09 \\
11667.27&8568.68&11700.95&8544.02 \\
11683.78&8556.57&11704.88&8541.15 \\
11683.92&8556.47&11706.39&8540.04 \\
11684.03&8556.39&11706.48&8539.98 \\
11684.18&8556.28&11706.55&8539.94 \\
11684.43&8556.10&11706.77&8539.77 \\
11684.58&8555.99&11709.79&8537.57 \\
11684.74&8555.87&11709.94&8537.46 \\
11684.95&8555.41&11710.29&8537.20 \\
11686.83&8554.34&11710.55&8537.01 \\
11687.14&8554.11&11711.17&8536.56 \\
11687.44&8553.89&11711.41&8536.38 \\
11687.72&8553.69&11711.53&8536.30 \\
11694.13&8549.00&11711.64&8536.22 \\
11694.39&8548.81&11713.72&8534.70 \\
11694.67&8548.61&11713.81&8534.64 \\
11697.42&8546.59&11720.67&8529.65 \\
\end{tabular}
\end{ruledtabular}
\end{table}

\setcounter{table}{0} 
\begin{table}
\caption{Wavenumbers and wavelengths of the detected CH$_3$I absorption lines (continued).}
\begin{ruledtabular}
\begin{tabular}{cccc}
$\nu'$ (cm$^{-1}$) & $\lambda$ (\AA) & $\nu'$ (cm$^{-1}$) & $\lambda$ (\AA) \\
\noalign{\smallskip} \hline \noalign{\smallskip}
11720.95&8529.44&11770.45&8493.57 \\
11730.12&8522.77&11772.70&8491.94 \\
11730.34&8522.61&11777.29&8488.63 \\
11730.43&8522.54&11777.94&8488.17 \\
11730.62&8522.41&11778.05&8488.09 \\
11738.96&8516.35&11778.27&8487.93 \\
11739.46&8515.99&11785.46&8482.75 \\
11739.58&8515.90&11785.65&8482.61 \\
11739.69&8515.82&11786.01&8482.35 \\
11739.80&8515.74&11788.60&8480.49 \\
11741.14&8514.77&11789.28&8480.00 \\
11741.39&8514.59&11798.00&8473.73 \\
11748.32&8509.57&11805.97&8468.01 \\
11761.04&8500.36&11806.13&8467.90 \\
11761.75&8499.85&11826.25&8453.49 \\
11767.15&8495.95&11827.97&8452.26 \\
11767.29&8495.85&11841.21&8442.81 \\
11767.42&8495.75&11841.39&8442.68 \\
11768.03&8495.31&11841.46&8442.63 \\
11768.13&8495.24&11842.02&8442.23 \\
\end{tabular}
\end{ruledtabular}
\end{table}

\begin{figure}
\resizebox{\columnwidth}{!}{
  \includegraphics{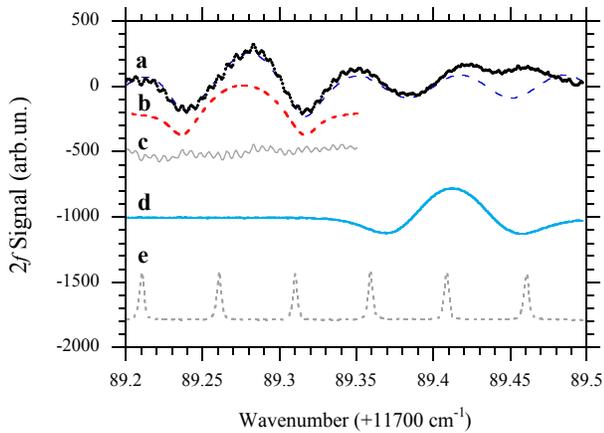}
}
\caption{2nd harmonic absorption signal of CH$_3$I around 848 nm (black dots) at $p = 31$ torr and RT, with the best fit (dashed blue line) ({\bf a}), the extracted peak fit ({\bf b)} and the residuals ({\bf c}), along with the H$_2$O reference signal ({\bf d}), all obtained by WMS with 10 Hz bandwidth. The F.-- P. interferometer transmission (f.s.r.= 0.05 cm$^{-1}$) is also shown ({\bf e}).}
\label{CH3I_H2O}
\end{figure}

Fig.~\ref{CH3I_H2O} presents the $2\!f$ absorption measurement signal for CH$_3$I at 11789.28 cm$^{-1}$, $T$ = 294 K, $p_{\mathrm{CH_3I}} = 31$ torr, and $m \simeq 2$, along with the best fit and its residuals. It displays also the H$_2$O $2\!f$ reference signal at 11789.41 cm$^{-1}$ and the transmission of the F.--P. interferometer  used for frequency linearization.
The evident etalon effect in the CH$_3$I measurement plot originates from reflections within the optical path, and it has been taken into account in the fit procedure. Only the higher frequency fringes remain in the fit residuals, originating from the confocal mirrors of the multipass cell; in fact the distance between them is $l = 42.8$ cm  and $\Delta\nu'  \cong$ 1$/(4l) = 0.0058$ cm$^{-1}$.

\begin{figure}
\resizebox{\columnwidth}{!}{
  \includegraphics{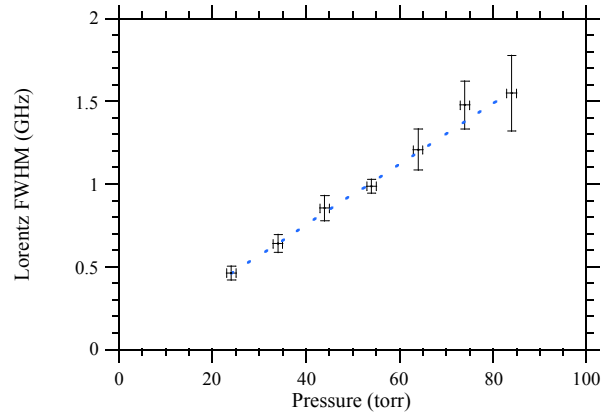}
}
\caption {Self-broadening measurements on the methyl iodide 11741.39 cm$^{-1}$ absorption line at RT obtained by TDLS and the procedure explained in the text. The blue dashed line shows the best linear fit.}
\label{CH3I_sb}
\end{figure}

The self-broadening coefficients ($\gamma_\mathrm{self}$) were measured for the first time for two CH$_3$I lines at RT and are shown in Table~\ref{tab:CH3ISB}.
\begin{table}
\caption{\label{tab:CH3ISB}Measured CH$_3$I HWHM self-broadening coefficients.}
\begin{ruledtabular}
\begin{tabular}{cc}
$\nu'$ (cm$^{-1}$) & $\gamma_\mathrm{self}$ (cm$^{-1}$/atm) \\
\noalign{\smallskip} \hline \noalign{\smallskip}
11741.39 & $ 0.23 \pm 0.02$ \\
11778.27 & $ 0.18 \pm 0.02$ \\
\end{tabular}
\end{ruledtabular}
\end{table}
\noindent  Fig.~\ref{CH3I_sb} shows the self-broadening measurement result for the line at 11741.39 cm$^{-1}$, where the Lorentzian component of the absorption line-width is plotted as a function of methyl iodide pressure at RT.

In the literature, we did not find any measurements of pressure line-broadening at these wavenumbers, but a comparison can be attempted with what was obtained in other spectral regions.

 In HITRAN database~\cite{ref:hitran2017} for the $\nu_4$ fundamental band at 3.3 $\mu$m the $\gamma_\mathrm{self}$ is reported between 0.1 a 0.5 cm$^{-1}$/atm at 296 K.
K.J. Hoffman et al.~\cite{ref:hoff2008} by diode laser absorption spectroscopy observed $\gamma_\mathrm{self}$ at RT for the $\nu_5$ band at 7 $\mu$m from 0.10 to 0.45 cm$^{-1}$/atm.
Raddaoui et al.~\cite{ref:radda2019} by a Fourier transform spectrometer at RT obtained $\gamma_\mathrm{self}$ ranging from 0.10 cm$^{-1}$/atm (at low and high quantum rotational parameter $J$) and 0.45 cm$^{-1}$/atm for the $\nu_6$ fundamental roto-vibrational band of CH$_3$I at around 11 $\mu$m.
Again, in this ro-vibrational ban at RT, Attafi et al.~\cite{ref:atta2019} obtained $\gamma_\mathrm{self}$ from 0.15 to 0.36 cm$^{-1}$/atm.
 Belli et al.~\cite{ref:belli2000} with the Doppler-free double-resonance technique have got the self collisional broadening parameters at RT for some absorption lines in the $\nu_6$ fundamental band, presenting values ranging from 0.18 to 0.20 cm$^{-1}$/atm.
Finally Ben Fathallah et al. using Fourier transform spectroscopy at RT obtained self-broadening coefficient values from 0.14 to 0.36 cm$^{-1}$/atm for the bands $\nu_5$ and $\nu_3 + \nu_6$~\cite{ref:fath2021}, while for the band $\nu_2$~\cite{ref:fathallah2021} at around 8 $\mu$m they obtained an average value of $\gamma_\mathrm{self}$ equal to 0.25 cm$^{-1}$/atm, with a minimum at 0.1 cm$^{-1}$/atm and a maximum at 0.4 cm$^{-1}$/atm.

Our results are within all of this ranges.

\section{Conclusion}
\label{conc}
82 new CH$_3$I absorption lines between 11660 and 11840 cm$^{-1}$ have been measured for the first time with a precision of 0.01 cm$^{-1}$, using a tunable diode laser spectrometer, the wavelength modulation spectroscopy, and the second harmonic detection in a 30 m Herriott-type multipass cell. A high modulation amplitude approach was adopted, and a dedicated fit function was developed. The strength of the observed lines varied between $10^{-27}$ and $10^{-26}$ cm/molecule at room temperature. For two of the observed lines, self-broadening coefficients similar to those reported for the same molecule in other spectral regions were obtained for the first time.
\section{Acknowledgments}
Thanks are due to A. Barbini for the electronic consultancy, to M. Tagliaferri and to M. Voliani for the technical assistance. The author is indebted to D. Bertolini for the calculations in the high modulation approximation.

\appendix*
\section{Frequency modulation in the high amplitude regime}
\label{appendix}

The use of high modulation amplitude $a$ is a necessity in order to increase the S/N ratio for very weak absortion lines.
In this case the derivative approximation of Eq.~(\ref{eq:comp}) no longer works and it is more appropriate to start from the other expression~\cite{ref:web1988}:
\begin{equation}
H_n(\nu,a) = \frac{2}{\pi} \int_0^\pi \tau(\nu + a \cos \theta)
\cos n\theta \; d\theta \, .
\label{eq:Hn}
\end{equation}

In order to extract the collisional component from the absorption line-shape, Arndt~\cite{ref:arn1965} and Wahlquist~\cite{ref:wahl1961} derived the analytic form of the harmonics for a Lorentzian function, which is the right choice when dealing with collisional broadening.

\noindent To do this they obtained the $n$th harmonic element inverting Eq.~(\ref{eq:mod}):
\begin{equation}
H_n(x,m) = \varepsilon_n \, \mathrm{i}^n \,
\int_{-\infty}^{+\infty} \hat{\tau}(\omega) \, J_n(m \omega) \,
e^{\mathrm{i} \omega x} \, d\omega
\label{inversa}
\end{equation}
where
\begin{equation}
\hat{\tau}(\omega) = \frac{1}{2\pi} \int \tau(x) \, e^{-\mathrm{i} \omega x} \, dx
\end{equation}
is the Fourier transform of the transmittance profile; $x=\nu/{\Gamma}$ and $m=a/{\Gamma}$, ${\Gamma}$ is the line-width; $J_n$ is the $n$th order Bessel function; $\varepsilon_0=1$, $\varepsilon_n=2$ ($n=1,2,\cdots$) and i is the imaginary unit.
The absortion cross-section in Eq.~(\ref{eq:absor}) is then put in the Lorentzian form:
\begin{equation}
\sigma_\mathrm{L}(x,m) \propto \frac{1}{1+(x+m \ cos\omega t)^2} \ \; .
\end{equation}
The second Fourier component of the absorption cross section can be recalculated following Arndt's work by setting $n=2$:
\begin{equation}
H_2(x,m) = - \frac{1}{m^2}
\biggl[\frac{\{[(1-\mathrm{i}x)^2+m^2]^{1/2}-(1-\mathrm{i}x)\}^2} {[(1-\mathrm{i}x)^2 + m^2]^{1/2}} + \mathrm{c.c.}\biggr]
\label{eq:imagi}
\end{equation}
and by eliminating the imaginary part:
\begin{widetext}
\beqn
H_2(x,m) = \frac{2}{m^2} - \frac{2^{1/2}}{m^2} \times \frac{1/2[(M^2+4x^2)^{1/2}+1-x^2][(M^2+4x^2)^{1/2} + M]^{1/2}+|x|\, [(M^2+4x^2)^{1/2}-M]^{1/2}}{(M^2+4x^2)^{1/2}}
\label{eq:sec}
\eeqn
\end{widetext}
where $$M=1-x^2+m^2 \, .$$
\begin{figure}{{\bf $2\!f$ line-shape}}
\resizebox{\columnwidth}{!}{
  \includegraphics{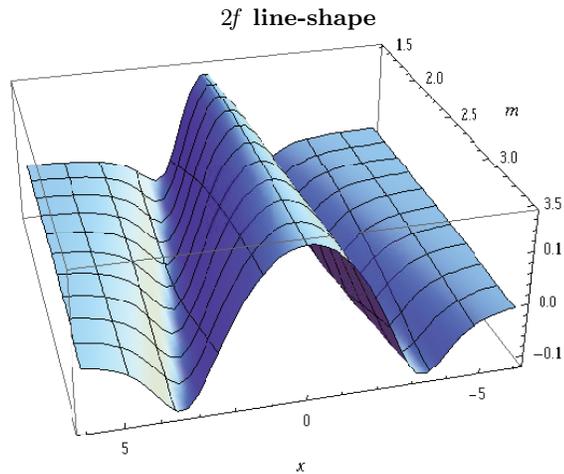}
}
\caption{Behavior of Eq.~(\ref{eq:sec}) as a function of the modulation parameter $m$.}
\label{modbrol}
\end{figure}

\noindent Eq.~(\ref{eq:sec}), close to the second derivative of the absorption feature only for low $m$, simulates the behavior of the line-shape at high modulation amplitudes.
This is shown in Fig.~\ref{modbrol}, where the equation is plotted in 3D, varying the modulation index $m$. For $m = 3$ the 2nd derivative is completely deformed by broadening, as it happens in reality.

\bibliography{BibCH3I.bib}

\end{document}